# Full C-Band WDM Transmission of Nonlinearity-Tolerant Probabilistically Shaped QAM over 2824-km Dispersion-Managed Fiber

Junho Cho[(1)]*, Xi Chen[(1)], Greg Raybon[(1)], Di Che[(1)], Ellsworth Burrows[(1)], and Robert Tkach [(1)]

[(1)] Nokia Bell Labs, Holmdel, NJ, USA, *junho.cho@nokia-bell-labs.com

**Abstract** *By tailoring probabilistic constellation shaping (PCS) for nonlinearity tolerance, we experimentally demonstrate up to 1.1 dB increase in signal-to-noise ratio (SNR) and 6.4% increase in total net data rate (NDR) compared to linear-channel-optimized PCS on a 2824-km dispersion-managed wavelength-division multiplexed (WDM) optical fiber link.*

## Introduction

Probabilistic constellation shaping (PCS) has significantly increased net data rate (NDR) of optical fiber communications recently. PCS is commonly optimized assuming additive white Gaussian noise (AWGN), where the theoretically maximum increase in NDR (known as the shaping gain) is achieved by infinite-length sphere shaping. However, in the presence of large fiber nonlinearities, such as in dispersion-managed links that still occupy a relatively large percentage of submarine links, the linearly-optimum PCS produces greater nonlinear interference than uniform signaling[1]–[6], resulting in a substantial reduction in shaping gain.

The potential advantage of finite-length shaping over infinite-length shaping in nonlinear fiber channels was first addressed in [1]. Since then there have been several approaches to mitigate the PCS-induced nonlinearities[2]–[5]. They showed that the statistical[2],[3] and temporal[4]–[6] properties of PCS change the signal-to-noise ratio (SNR) by changing the nonlinear signal interaction. In particular, it was shown by simulations[5],[6] and experiments[4] that the shorter the sphere shaping block length, the larger the SNR. However, none of the existing works investigated the impact of symbol rates (i.e., *spectral* properties) of PCS on nonlinearities, and the experiments were limited to single-channel transmission. Although the relation between symbol rates and nonlinearities was studied for uniform signaling[7], the results cannot be applied to finite-length PCS because in PCS, unlike in uniform signaling, strong temporal correlation exists between symbols.

In this work, we show the influence of the temporal and spectral properties of PCS on nonlinear transmission performance, and demonstrate up to 1.1 dB increase in SNR (displayed in Fig. 3 below) and 6.4% increase in total NDR in a 3.7-THz-wide full C-band transmission system with 2824-km dispersion-managed optical fiber, by jointly adapting the sphere shaping block length $n$ and symbol rate $R_{Sym}$ for each wavelength-division multiplexed (WDM) channel.

## Experimental Setup

As shown in Fig. 1(a), our experimental recirculation loop consists of 7 spans of 40.3 km (on average) fiber, 1 span of which (span #4) is standard single-mode fiber (17.24 ps/nm/km dispersion, with slope of 0.092 ps/nm$^2$/km) and the rest are negative-dispersion fiber (-2.47 ps/nm/km dispersion on average, with slope of -0.1026 ps/nm$^2$/km, all measured at 1550.13 nm). We transmit 37 100-GHz-wide WDM channels in the C-band over 10 loops (2824 km). Fig. 1(b) shows the accumulated dispersion as a function of distance for several selected channels. We denote the channels at 192.1 to 195.7 THz sequentially by Ch#1 to Ch#37, then Ch#8 at 192.8 THz (solid red line) undergoes zero net dispersion, and the farther away from Ch#8 the channel undergoes greater net dispersion.

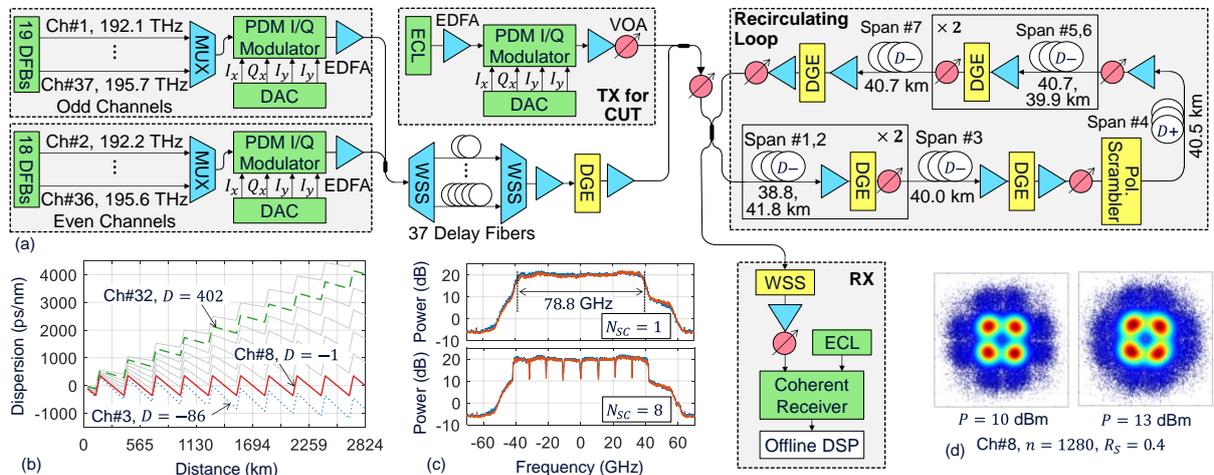

**Fig. 1**: (a) Experimental setup, (b) dispersion map, (c) received power spectral density, and (d) recovered constellation at Ch#8.
* DFB: distributed feedback laser, DAC: digital-to-analog converter, ECL: external cavity laser, WSS: wavelength selective switch

We transmit three independent sets of modulation patterns, one of which is loaded on the channel under test (CUT), the other two on the even- and odd-indexed interfering channels that are fully decorrelated by delay fibers of distinct lengths ($10, 20, \ldots, 190$ m for odd channels, and $10, 20, \ldots, 180$ m for even channels, cf. Fig. 1(a)). When evaluating the CUT, all channels are loaded with a PCS format of the same shaping rate $R_S$ (bits per positive amplitude), same shaping block length $n$ (amplitudes), and same symbol rate $R_{Sym}$ (GBd) as the CUT. At the receiver (RX), commonly used coherent digital signal processing (DSP) is performed offline to recover the signals.

**Optimization Strategy**

We adjust $n$ and $R_{Sym}$ (hence exploring both the *temporal* and *spectral* properties of PCS) to maximize the NDR. First, at each $R_{Sym}$, we vary $n \in \{20, 40, 80, 320, 1280\}$ (temporal duration of a shaped signal block). We realize sphere shaping using the enumerative sphere shaping (ESS)[8] for $n = 20, 40, 80, 320$ and constant composition distribution matching (CCDM)[9] for $n = 1280$, and embed it in the probabilistic amplitude shaping (PAS) architecture[10]. Fig. 2 shows the (linear) shaping gap of the realized sphere shaping for selected $R_S$, as quantified by the increase in average energy relative to ideal Maxwell-Boltzmann (MB)-distributed amplitudes of the same $R_S$. We map 4 consecutive shaped amplitudes onto a single polarization-division multiplexed (PDM) symbol for enhanced nonlinearity tolerance[6]. Second, for each $n$, we vary $R_{Sym}$ (hence the *spectral* bandwidth of a shaped signal block) by performing digital sub-carrier multiplexing (DSM) on each channel that occupies ~80 GHz. We use $N_{SC} \in \{1,2,4,8\}$ digital subcarriers to emulate symbol rates $R_{sym} \in \{78.8, 39.4, 19.7, 9.8\}$ GBd. The same $R_S$ is loaded on all subcarriers. Root-raised cosine (RRC) pulse shaping is performed on each subcarrier with a roll-off factor flexibly adjusted between 0.05 and 0.1 depending on $N_{SC}$ (cf. Fig. 1(c)). This is to circumvent the bandwidth limitations of the transmitter favoring $N_{SC} = 1$. With this, we ensure in a back-to-back experiment that the SNR averaged over subcarriers differs by no more than 0.2 dB between DSM of all $N_{SC}$ at the same optical SNR (OSNR). By noting that sphere shaping limits the total energy of a block of signals to within a fixed value, and since each shaped block occupies $n/4$ symbols in time (division by 4 due to PDM) and $R_{Sym}$-GHz-wide band in frequency (neglecting RRC filtering), our strategy finds the sphere shaping block size that optimally limits the total signal energy in time ($n/4$) and frequency ($R_{Sym}$) such that the nonlinearities in fiber are minimized.

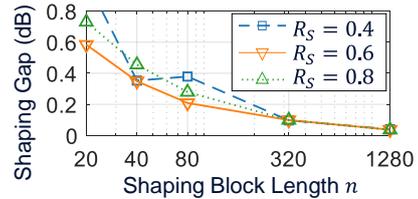

**Fig. 2**: Shaping gap of finite-length sphere shaping.

For each $n$ and $R_{Sym}$, the achieved NDR is determined by finding the maximum shaping rate $R_S^* \in \{0.3, 0.4, \ldots, \}$ that yields a normalized generalized mutual information (NGMI)[11] greater than the threshold $NGMI^* = 0.861$ of a field-programmable gate array (FPGA)-verified spatially-coupled low-density parity-check (LDPC) code with rate $R_C = 0.8$[12]. The NDR per optical carrier is then given by
$$NDR = \alpha\{R_S^* + [1 - m(1 - R_C)]\}R_{Sym}N_{SC}R_{Pi} \quad (1)$$
in Gb/s, where $\alpha = 4$ accounts for PDM, $m = \log_2 M$ for $M^2$-ary PCS quadrature amplitude modulation (QAM), and $R_{Pi} = 47/48$ is the pilot ratio used in this work. For PCS 16-QAM with $R_S = 0.2$, quadrature phase-shift keying (QPSK) is used instead, since they produce the same NDR.

**Experimental Results: Nonlinearity Mitigation by Short Shaping and High Symbol Rates**

Throughout the paper, we use $n = 1280$ as the benchmark for comparison with shorter sphere shaping since it approaches ideal MB shaping to within 0.05 dB for every $R_S$ under test (cf. Fig. 2). An intermediate symbol rate of $R_{Sym} = 39.4$ GBd is chosen for the benchmark, which will be shown to be optimum for high-dispersion channels with least nonlinearities.

To find the optimal $n$ and $R_{Sym}$, the SNR is evaluated for selected 8 channels across the C-band. Here, the SNR is a better figure of merit than the NDR to compare PCS in terms of nonlinearity mitigation. This is because the SNR is changed almost solely by the PCS-induced nonlinear noise, whereas the NDR is changed by both the nonlinear noise and the (linear) shaping gap. We also vary the launch power $P$ (the total power for all channels) from 8 to 13 dBm in 1-dBm steps, with the power excursion across the C-band being maintained within 4 dB after 10 loops by adjusting the inline dynamic gain equalizers (DGEs, cf. Fig. 1(a)). Typical recovered constellations are shown in Fig. 1(d).

Fig. 3(a) shows the exemplary SNRs measured as a function of $n$ at three different dispersions $D = -86, -1, 402$ ps/nm/loop (at Ch#3, #8, #32, cf. Fig. 1(b)). With large $D = 402$ (triangles), SNR does not change much with $n$, regardless of $P$. On the other hand, with small $D = -1$ (circles) at high $P = 13$ dBm (solid lines), SNR generally improves as $n$ decreases. This shows that short sphere shaping makes nonlinear interference smaller than long sphere shaping. Compared to $n = 1280$, $n = 20$ improves SNR by 1.1 dB (stars in Figs. 3(a) and (c)). Note however that the large shaping gap of $n = 20$ (cf. Fig. 2) also plays an important role in determining optimal $n^*$, as will be shown below.

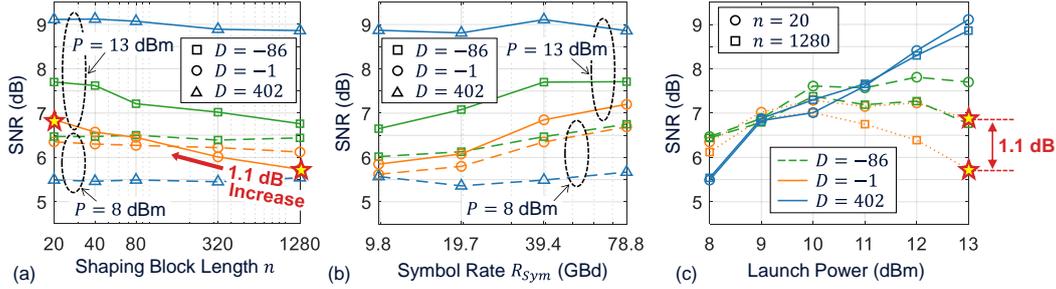

**Fig. 3**: SNR drawn as a function of (a) the block length $n$ at $R_{Sym} = 39.4$ GBd (i.e., $N_{SC} = 2$), (b) the symbol rate $R_{Sym}$ with $n = 1280$, and (c) the launch power $P$ at $R_{Sym} = 39.4$ GBd with $n = 20, 1280$.

Fig. 3(b) shows the SNR as a function of $R_{Sym}$. With large $D = 402$ (triangles) at high $P = 13$ dBm (solid lines), SNR is highest at an intermediate symbol rate of $R_{Sym} = 39.4$ GBd (hence used for the benchmark), although the difference in SNR between various $R_{Sym}$ is small. On the other hand, with small $D = -1$ (circles), SNR generally improves as $R_{Sym}$ increases, up to the tested maximum $R_{Sym} = 78.8$ GBd. The SNR improvement reaches 0.4 dB and 1.4 dB compared to $R_{Sym} = 39.4$ and 9.8 GBd, respectively.

**Experimental Results: NDR Maximization**
Knowing the various dependencies as above, we now maximize the NDR across *all* the 37 WDM channels by adapting $n$ and $R_{Sym}$ individually for each channel, where NDR is determined by Eq. (1). We use the launch powers of $P = 12$ and 13 dBm that are higher than the optimum in low-dispersion channels but suitable for high-dispersion channels, cf. Fig. 3(c). Note that these launch powers are optimum for QPSK even at 192.8 THz with zero net dispersion. The figures below show the results of $P = 13$ dBm.

Fig. 4(a) shows the NDR-maximizing $n^*$ for all channels. In general, $n^*$ decreases as the dispersion $|D|$ (green solid line) decreases. However, the shortest $n = 20$ is not optimal for any channel since its large shaping gap (cf. Fig. 2) outweighs the SNR improvement. Fig. 4(b) shows that transition between $R^*_{Sym} = 78.8$ GBd and 39.4 GBd occurs at $D = 200$ to 400. Using the $n^*$ and $R^*_{Sym}$ of Figs. 4(a) and (b), the SNR of Fig. 4(c) (circles) is obtained. The SNR of the benchmark PCS is also shown (triangles), with which the NGMI condition is not fulfilled by any shaping rate at the zero-dispersion regime (192.8 THz), hence the channels are modulated with QPSK (pluses in Fig. 4(c)).

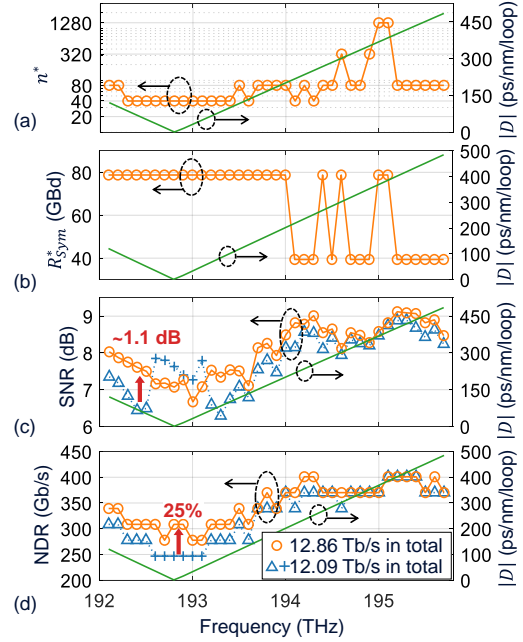

**Fig. 4**: (a) Optimal shaping block length $n^*$, (b) optimal symbol rate $R^*_{Sym}$. (c) SNR and (d) NDR, achieved by the optimized $n^*$ and $R^*_{Sym}$ (circles) and by the benchmark PCS with $n = 1280$ at $R_{Sym} = 39.4$ GBd (triangles, pluses).

Fig. 4(d) shows the NDRs achieved by $n^*$ and $R^*_{Sym}$ (circles), and by the benchmark (triangles), where the optimized PCS produces higher NDRs than the benchmark widely in low-dispersion channels. The *total* NDR increase in the C-band by the nonlinearity-optimized PCS reaches 6.4 % (12.86 Tb/s as compared to 12.09 Tb/s of the benchmark, obtained as the sum of the NDRs of all 37 channels in Fig. 4(d)) at 13 dBm. The total NDR increase at 12 dBm is also 6.4% (13.29 Tb/s compared to 12.49 Tb/s). At 13 dBm, the NDR increase reaches 25% in the worst-performing channels in the zero-dispersion regime (cf. red arrow in Fig. 4(d)) and 12.1% on average over the underperforming 20 channels with low dispersion.

**Conclusion**
We demonstrated through extensive WDM experiments that PCS-induced nonlinearities can be minimized by adjusting $n$ and $R_{Sym}$ for each channel, and achieved increases of 12.1% and 6.4% in NDR, respectively, in the underperforming nonlinear channels and in the entire system.

## References

[1] R. Dar et al., "On shaping gain in the nonlinear fiber-optic channel," in *Proc. IEEE Int. Symp. Inf. Theory*, Honolulu, HI, USA, Jun. 2014, pp. 2794–2798.

[2] J. Cho et al., "Low-complexity shaping for enhanced nonlinearity tolerance," in *Proc. Eur. Conf. Opt. Commun.*, Dusseldorf, Germany, Sep. 2016, Paper W1C2.

[3] T. Fehenberger et al., "On probabilistic shaping of quadrature amplitude modulation for the nonlinear fiber channel," *J. Lightw. Technol.*, vol. 34, no. 21, pp. 5063–5073, Nov. 2016.

[4] S. Goossens et al., "First experimental demonstration of probabilistic enumerative sphere shaping in optical fiber communications," in *Proc. Opto-Electron. Commun. Conf.*, Fukuoka, Japan, Jul. 2019, Paper PDP 2.

[5] A. Amari et. al., "Introducing enumerative sphere shaping for optical communication systems with short blocklengths," *J. Lightw. Technol.*, vol. 37, no. 23, pp. 5926–5936, Dec. 2019.

[6] T. Fehenberger et. al., "Analysis of nonlinear fiber interactions for finite-length constant-composition sequences," *J. Lightw. Technol.*, vol. 38, no. 2, pp. 457–465, Jan. 2020.

[7] P. Poggiolini et al., "Analytical and experimental results on system maximum reach increase through symbol rate optimization," *J. Lightw. Technol.*, vol. 34, no. 8, pp. 1872–1885, Apr. 2016.

[8] F. M. J. Willems and J. J. Wuijts, "A pragmatic approach to shaped coded modulation," in *Proc. Symp. Commun. Veh. Technol.*, Oct. 1993, pp. 1–6.

[9] P. Schulte and G. Böcherer, "Constant composition distribution matching," *IEEE Trans. Inf. Theory*, vol. 62, no. 1, pp. 430–434, Jan. 2016.

[10] G. Böcherer F. Steiner, and P. Schulte., "Bandwidth efficient and rate-matched low-density parity-check coded modulation," *IEEE Trans. Commun.*, vol. 63, no. 12, pp. 4651–4665, Dec. 2015.

[11] J. Cho, L. Schmalen, and P. Winzer, "Normalized generalized mutual information as a forward error correction threshold for probabilistically shaped QAM," in *Proc. Eur. Conf. Opt. Commun.*, Gothenburg, Sweden, Sep. 2017, Paper M.2.D.2.

[12] L. Schmalen et al., "Spatially coupled soft-decision error correction for future lightwave systems," *J. Lightw. Technol.*, vol. 33, no. 5, pp. 1109–1116, Mar. 2015.